\documentstyle[namedreferences,psfig]{spse2006}

\renewcommand{\thepage}{}   

\renewcommand{\thepage}{}   

\runningtitle{Search for solar neutrino radiative decays}  
\begin{opening}
  \title{Search for possible solar neutrino radiative decays during total solar eclipses}       
  \author{S. Cecchini\thanks{Also IASF/CNR, I-40129 Bologna, Italy}}
  \author{D. Centomo}
  \author{G. Giacomelli}
  \author{R. Giacomelli}
  \author{V. Popa\thanks{Also Institute for Space Sciences, R-77125 Bucharest 
  M\u{a}gurele, Romania}\thanks{presenting author}}
  \institute{Dipartimento di Fisica dell'Universit\`{a} and INFN Sezione di 
  Bologna, I-40127 Bologna, Italy}

\end{opening}

\begin{document}
\begin{abstract}
  Total solar eclipses (TSEs) offer a good opportunity to look for photons produced in
  possible radiative decays of solar neutrinos. In this paper we briefly review the
  physics bases of such searches as well as the existing limits for the 
  $\nu_2$ and $\nu_3$ proper lifetimes obtained by such experiments. We then report
  on the observations performed in occasion of the 29 March 2006 TSE, from Waw an 
  Namos, Libya.  

\end{abstract}

\section{Introduction}
In the last years, the evidence of neutrino (both solar and atmospheric) oscillations clearly 
shown that neutrinos have non-vanishing masses, and that the neutrino flavor
eigenstates ($\nu_e$, $\nu_\mu$ and $\nu_\tau$) are superpositions of mass eigenstates 
($\nu_1$, $\nu_2$ and $\nu_3$) \cite{macro,sk,sno}. Neutrinos could undergo radiative decays,
e.g. $\nu_2 \rightarrow \nu_1 + \gamma$, as initially suggested in \cite{melott81}; the present 
status of the neutrino decay theory is summarized in \cite{sciama95}. 

The neutrino radiative decay requires a non-vanishing neutrino magnetic moment; very stringent
existing limits ($\mu_\nu < 1.3 \times 10^{-10} \mu_B$, \cite{hagiwara02}) refer to neutrino 
flavor 
eigenstates and thus are not directly applicable to possible dipole magnetic moments of neutrino 
mass eigenstates.

Neutrino decays (radiative or not) were searched for from astrophysical phenomena
 such as Supernova physics or the absence of $\gamma$ rays in the Sun radiation, or from 
infra-red background measurements. Such neutrino lifetime lower limits are typically very 
large (e.g $\tau_0/m > 2.8 \times 10^{15}$ s/eV where $\tau_0$ is the lower proper 
lifetime limit for a neutrino of mass $m$, 
\cite{bludman92}), but they are indirect and rather speculative. 

Much lower  ``semi-indirect" limits were obtained from the re-interpretation of solar and 
atmospheric neutrino data. Although the present accepted interpretation of the existing 
observations is essentially that of neutrino oscillations, the hypothesis of neutrino decays 
cannot be completely discarded, as a secondary effect. As an example, from the SNO solar 
neutrino data, in \cite{bandy03} a lower limit of 
$\tau_0/m > 8.7 \times 10^{-5}$ s/eV was inferred. By combining all available solar neutrino 
data, limits as $\tau_0/m > 2.27 \times 10^{-5}$ s/eV \cite{jos02}, or, following different 
assumptions, $\tau_0/m > 10 \times 10^{-4}$ se/V \cite{beacon02} were obtained.
  
Direct searches for radiative neutrino decays (not correlated with TSEs) were also performed: 
in the vicinity of nuclear reactors (e.g. \cite{bouchez88}, yielding 
limits between $\tau_0/m > 10^{-8}$ s/eV and $\tau_0/m > 0.1$ s/eV, for $\Delta m / m$ between 
$10^{-7}$ and 0.1), or using the Borexino Counting Test Facility at Gran Sasso, at the level of
  $\tau_0/m \simeq 10^3$  s/eV, as a function of the neutrino polarization \cite{derbin02}. 

The Sun is a very strong source of $\nu_e$ neutrinos; the expected flux at the 
Earth (neglecting oscillation effects) is $\Phi \simeq 7 \times 10^{10}$ cm$^{-2}$s$^{-1}$.
In normal conditions, if radiative neutrino decays occur yielding visible photons, they would not be 
observable due to the very large amount of light produced by the Sun. During a TSE, the Moon 
absorbs the direct light from the Sun, but is completely transparent to solar neutrinos. 
An experiment looking for such an effect is thus sensitive to neutrino decays occurring in the 
space between the Moon and the Earth. 

In a pioneering experiment performed in occasion of the October 24, 1995 TSE, a first search 
was made for visible photons emitted through radiative decays of solar neutrinos during  their 
flight between the Moon and the Earth \cite{birnbaum97}. The authors assumed that all neutrinos 
have masses of the order of few eV, $\delta m^2 \simeq 10^{-5}$ eV$^2$, and an average energy of
 860 keV. Furthermore, they assumed that all decays would lead to visible photons, which would 
travel nearly in the same direction as the parent neutrinos. In the absence of a positive signal,
 the search yielded a lower proper lifetime of 97 s. 

In \cite{cecco04a} we discussed a complete Monte Carlo model for the solar neutrino radiative 
decay. The neutrino production inside the core of the Sun is described according to the 
Standard Solar Model (SSM) predictions \cite{bahcall98}, both from the point of view of the 
energy spectrum and of the source geometry. We considered the $\nu_1$ mass $m_1$ between
$10^{-3}$ and 0.3 eV, and $\nu_1$ the lowest mass neutrino state. The mass differences 
$\delta m^2 = m_2^2 - m_1^2= 6 \times 10^{-5}$ eV$^2$ and $\Delta m^2 =
m_3^2 - m_2^2 \simeq m_3^2 - m_1^2 = 2.5 \times 10^{-3}$ eV$^2$ are chosen in agreement 
with the solar and atmospheric neutrino oscillation experimental data, respectively.
 The mixing angles were also considered as measured by oscillation experiments, and the still 
unknown $\theta_{13}$ was chosen as $\sin^2 \theta_{13} \simeq 0.1$. In calculating the zenith 
angular distributions of the emitted photons we 
considered the probability density functions (pdfs) (corresponding to 3 different neutrino 
polarizations: $\alpha = -1$ (lefthanded neutrinos), $\alpha =0$ (Majorana neutrinos) and 
$\alpha = 1$ (righthanded neutrinos). The pdfs were integrated according to the angular 
resolution of the simulated experiments. The simulations yielded narrow visible signals 
corresponding to the solar $\nu_2 \rightarrow \nu_1 + \gamma$ decays, concentrated in 
about 100 arcsec in the direction of the center of the Sun. The signal corresponding to 
$\nu_3 \rightarrow \nu_{1,2} + \gamma$ decays is expected to be broader, with annular maxima 
at about 250 arcsec away from the direction pointing to the center of the Sun.

We made our first search for solar neutrino radiative decays during the June 21, 2002 TSE, 
in Zambia \cite{noi}. Two data sets were obtained and analyzed: 4149 frames recorded by a 
digital video-camera equipped with a 2$\times$ telelens and a 10$\times$  optical zoom, and 10 
digital photographs made with a Matsukov-Cassegrain telescope (90 mm $\Phi$ and $f = 1250$ mm).
 The proper lower time limits (95\% C.L.) obtained for the $\nu_2 \rightarrow \nu_1 + \gamma$ 
decays of lefthanded neutrinos range from $\tau_0/m_2 \simeq 10$ to $\simeq 10^9$ se/V, 
for $10^{-3}$ eV $ < m_1 < 0.1$ eV. These limits are among the best obtained from direct 
measurements, demonstrating the potentiality of neutrino decay experiments during TSEs. 

In this paper we report on the observations performed in occasion of the  29 March 2006 TSE, 
from Waw an Namos, Libya. We describe the experimental set-up and make a first characterization 
of the obtained data. The data analysis in terms of solar neutrino radiative decays is an 
ongoing lengthy process, so we cannot give here estimates on the lifetimes or lower limits. 

\section{29 March 2006 total solar eclipse observation}     

Considering the good quality of transport and on-site facilities provided by the organizers of 
the Eclipse event in Libya, we prepared a main experiment designed to reach a sensitivity 
improvement of at least three orders of magnitude relative to our 2002 observations \cite{noi}. 
For redundancy, smaller back-up experiments were also prepared.

\subsection{The main experiment}

\begin{figure}[htbp]
  \centerline{\psfig{figure=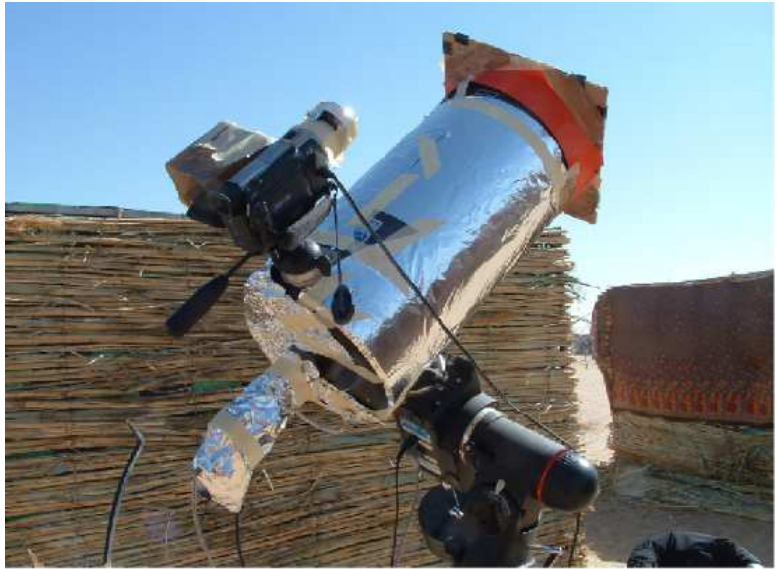,width=0.8\linewidth}}    
  \caption[]{\label{notte}
  The main experiment installed on the observation site, in the Waw an Namos eclipse camp. See text for details. }  
\end{figure}

Our main experiment used a Matsukov-Cassegrain telescope ($\Phi=235$ mm, $f = 2350$ mm), 
equipped with a fast Mx916 CCD camera. The original findscope was substituted by the same digital
 videocamera that we used for data taking during the 2002 TSE. Figure \ref{notte} shows our 
apparatus, mounted at the observation site; the solar filter was removed during the totality 
phase. 

The night before the TSE we aligned the system, adjusted the focus and took  calibration images 
of some standard luminosity stars. In order to avoid the over-heating of the telescope and CCD 
during the eclipse day and to minimize the possibility of focus and alignment changes, the 
equipment was protected with aluminum foils. 

The telescope movement was set to follow the Sun, in order to have always the center of the 
acquired images coincident with the Sun center. Furthermore, we implemented a special CCD 
exposure algorithm, in order to adapt the  exposition times to the luminosity level of the
Moon image. Although the ashen light is one of the main background sources in the search for a 
 signal produced by solar neutrino radiative decays, it allows the reconstruction, 
frame by frame, of the real position of the Sun behind the Moon, eliminating the risk of 
pointing errors due to undesired movements of the telescope tripod on the sand.
The algorithm analyzed in real time the luminosity of the previous registered frames, and 
determined the exposure time for the next frame for an average luminosity of the image at half 
value of the CCD dynamical range, and maximizing the contrast.

\begin{figure}[htbp]
  \centerline{\psfig{figure=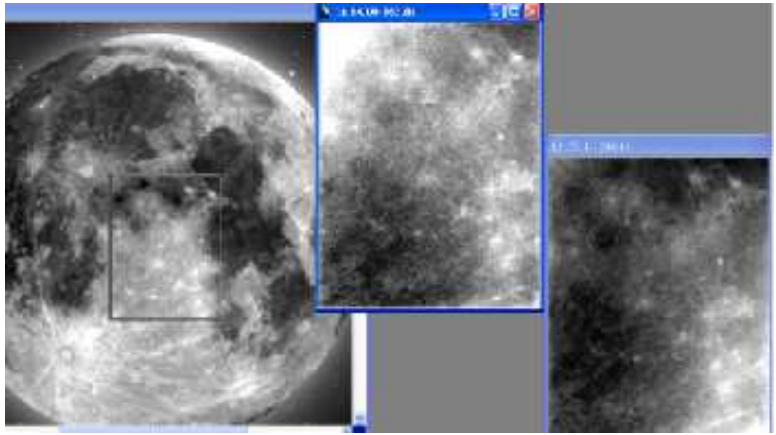,width=0.8\linewidth}}    
  \caption[]{\label{luna}
  Two frames obtained by our experiment at the beginning of the totality phase (middle) and near to its end (right), compared with an image of the full Moon (left). The approximate field of view of our observations is also shown on the full Moon picture}  
\end{figure}

In Fig. \ref{luna} we present a comparison between a full Moon image (left panel) and two frames
 registered by our experiment at the beginning of the totality phase (middle panel) and near to 
its end (right panel). The approximate field of view of our observations is also marked by the 
rectangle on the full Moon picture.
  
\begin{figure}[htbp]
  \centerline{\psfig{figure=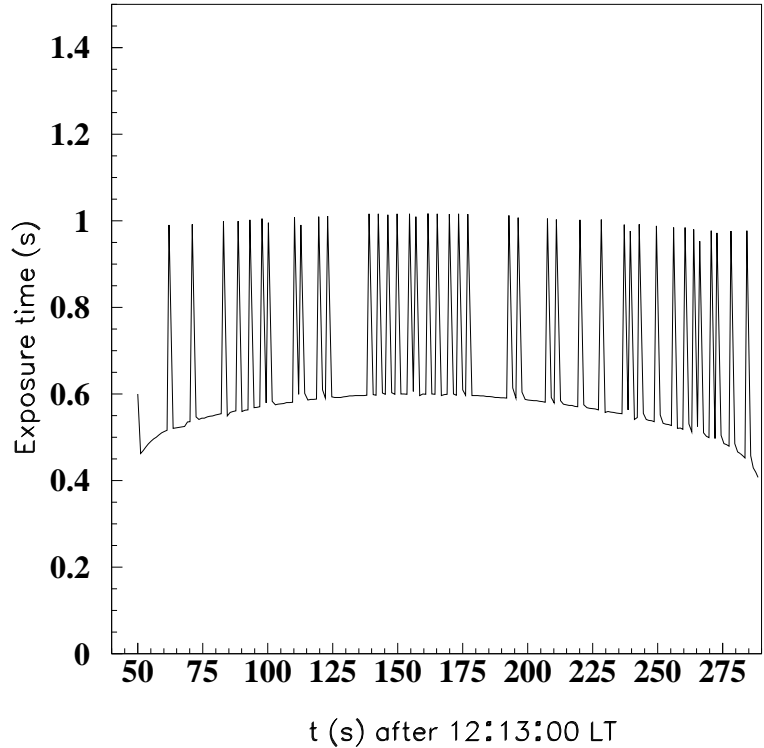,width=0.9\linewidth}}    
  \caption[]{\label{figtimp}
 The distribution of the CCD exposure time, during the totality phase of the eclipse.}  
\end{figure}

The luminosity change between frames may come from  
 diffraction of the coronal light in the Earth's atmosphere.
 The effect was taken into consideration by our automatic exposure algorithm, as shown in 
Fig. \ref{figtimp}.

For most of the frames, the exposure time was about 0.5 s; some attempts were made by the 
algorithm to increase the exposure time to about 1 s, all of them after a frame that had a 
lower luminosity than the required value. This could also be due to some fluctuations in the 
electric power affecting the acquisition, but since we have all the information concerning the 
particular conditions in which each frame was recorded, this will not interfere on further 
analyses. 

\begin{figure}[htbp]
  \centerline{\psfig{figure=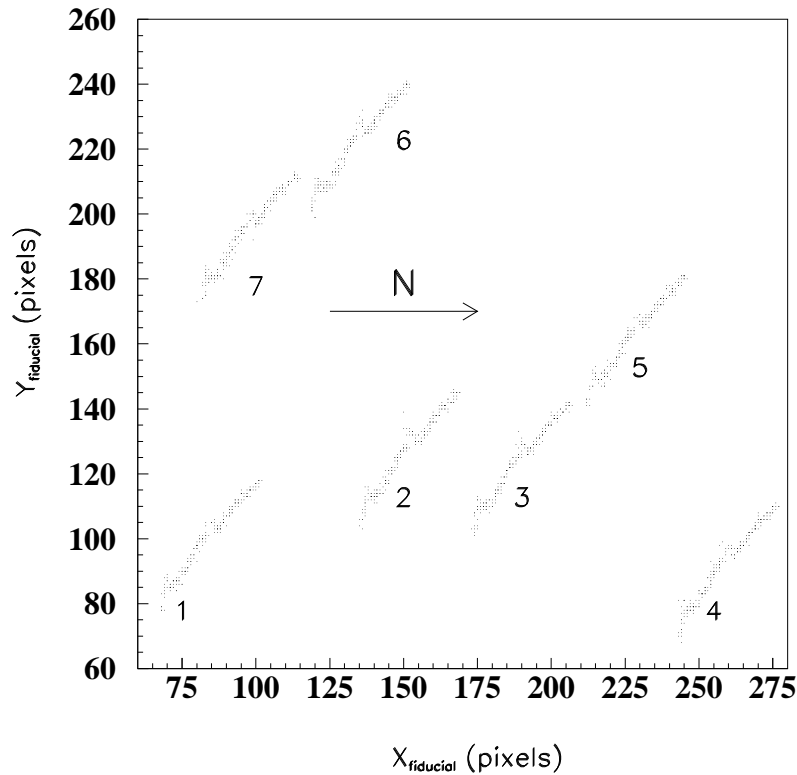,width=0.9\linewidth}}    
  \caption[]{\label{figmap}
 The movement of the selected ``fiducial points" (small luminous craters on the Moon surface) 
during the totality phase of the eclipse. The arrow indicates the North direction.}  
\end{figure}

In all panels of Fig. \ref{luna} a number of small luminous craters is easily observable. We 
selected 7 of them, to be used as ``fiducial points" in order to test the telescope movements 
and to reconstruct the position of the Sun for each of the about 200 frames taken. The 
displacement of those markers during the totality phase is shown in Fig. \ref{figmap}; note that
 the North direction is from left to right, due to the CCD relative position.
  
From Fig. \ref{figmap} it is clear that the fiducial points represent the Moon's movement 
in front of the eclipsed Sun, with some deviations in some frames taken in the middle and at 
the end of the totality. Those small pointing jumps were probably due to human activity in the 
vicinity of the telescope, but they may be corrected using the actual coordinates of the 
reference craters. 

\subsection{The backup experiments}

The digital video-camera (the same as that used in occasion of the 2002 TSE) was itself a 
small backup experiment. It produced a digital film of the TSE, that could at least confirm our 
earlier results \cite{noi}.

\begin{figure}[htbp]
  \centerline{\psfig{figure=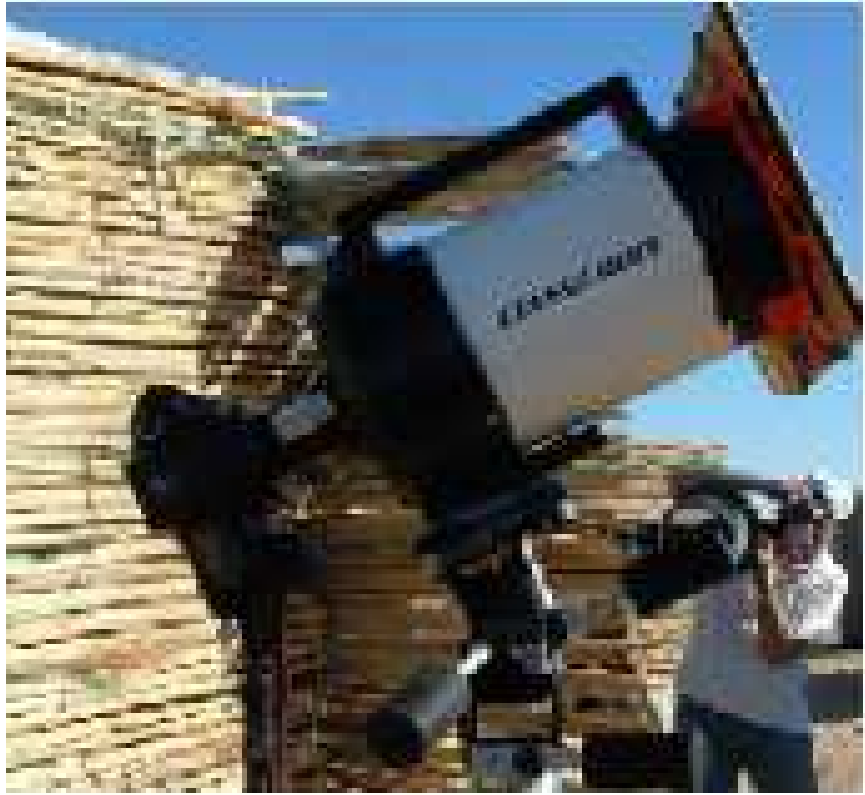,width=0.9\linewidth}}    
  \caption[]{\label{mic}
 The backup telescope with the digital camera, on the observation site.}  
\end{figure}

\begin{figure}[htbp]
  \centerline{\psfig{figure=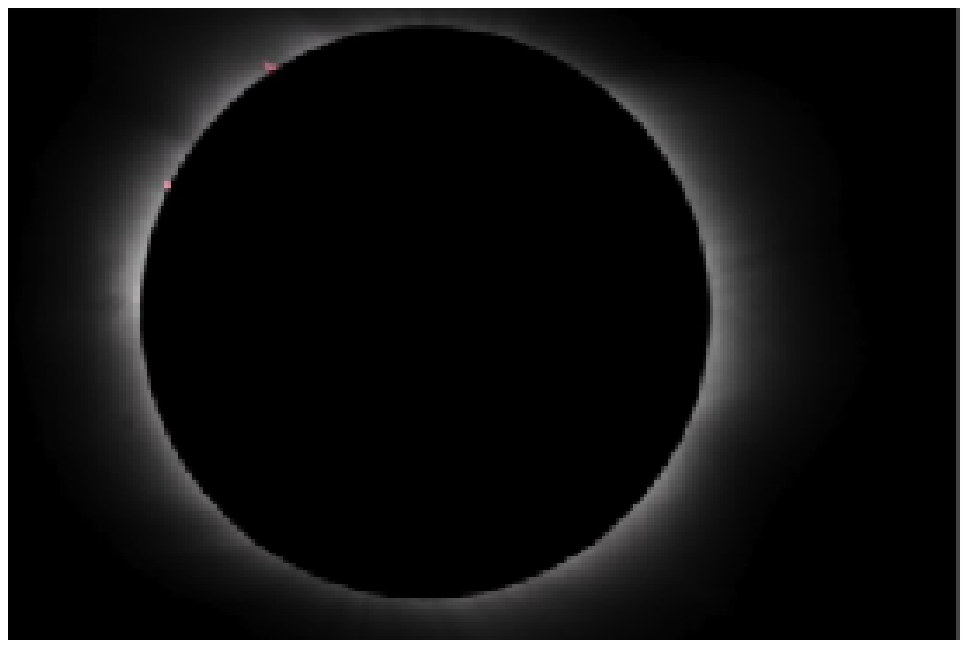,width=0.9\linewidth}}    
  \caption[]{\label{cecco}
 One of the 50 digital photographs obtained with the backup experiment.}  
\end{figure}

We also used a smaller Celestron C5 telescope, equipped with a manually controlled digital 
camera (Canon D20), as shown in Fig. \ref{mic}.
We obtained 50 digital pictures of the eclipse, as the one presented in Fig. \ref{cecco}; they will be analyzed in the same way as reported in \cite{noi}.
All those pictures are available from the science popularization site 
http://www.scienzagiovane.unibo.it.  

\section{Conclusions}

In occasion of the March 29 2006 total solar eclipse we made several observations from the 
Waw an Namos camp, in southern Libya, looking for a visible light signal possibly produced in 
solar neutrino radiative decays, during the neutrino flight between the Moon and the Earth.

Our main data consist of a set of about 200 CCD images of the central area of the Moon, 
aligned with respect to the position of the center of the Sun. 
 The performant instrumentation used and the specially developed algorithm optimizing the 
exposure time allowed to  visualize on each frame clearly recognizable details of the Moon 
surface, in the light reflected by the Earth. This will allow a precise reconstruction of the 
relative position of the Sun and Moon in each frame,
 thus eliminating the pointing errors due by accidental movements in the vicinity of the 
apparatus. Furthermore, as the ashen light is one of the major background sources in our search,
 the quality of the obtained images suggests that we can reach an experimental sensitivity close
 to the best possible in measurements using the visible spectrum.

The backup experiments worked also very well: we have 50 digital photographs of the total solar
 eclipse, obtained with a smaller telescope and digital films obtained with two digital 
videorecorders, that may also be used for pedagogical puposes.

The main data are currently under analysis; if no positive signal will be found, we expect a 
major improvement of our previous limits \cite{noi}. They will also allow a precise measurement of the Earth's albedo. 

\vspace{3ex} \footnotesize \noindent 
{\bf Acknowledgments.}
We are grateful to the organizers of the SPSE 2006 event, for all their efforts that allowed
to perform our experiment and to join the very exciting Symposium in Waw an Namos. 
Special thanks are due to the Winzrik Group and to the Libyan Air Force, for their kind and 
efficient assistance. 


\end{document}